\documentclass[12pt,letter]{article}
\pdfoutput=1
\usepackage{graphicx, epsfig, color,cite}
\usepackage{amsmath,amsthm,amsfonts,geometry,etoolbox}

\usepackage{amssymb}
\usepackage{caption,subcaption,graphicx,appendix}
\usepackage[hidelinks]{hyperref}
\def\bi{\begin{itemize}}
\def\ei{\end{itemize}}

\def\tst{\tilde t}

\def\tg{\tilde g}

\def\alt{\lesssim}
\def\agt{\gtrsim}
\def\be{\begin{equation}}  
\def\ee{\end{equation}}  
\def\bea{\begin{eqnarray}}  
\def\eea{\end{eqnarray}}

\newcommand{\myeq}{\begin{small}\begin{equation}\begin{aligned}}
\newcommand{\myeqend}{\end{aligned}\end{equation}\end{small}}

\begin{document}
\begin{titlepage}
\begin{flushright}
OU-HEP-220702
\end{flushright}

\vspace{0.5cm}
\begin{center}
  {\Large \bf Fine-tuned vs. natural supersymmetry:\\
    what does the string landscape predict?}\\

\vspace{1.2cm} \renewcommand{\thefootnote}{\fnsymbol{footnote}}
{\large Howard Baer$^{1}$\footnote[1]{Email: baer@ou.edu },
Vernon Barger$^2$\footnote[2]{Email: barger@pheno.wisc.edu},
Dakotah Martinez$^1$\footnote[4]{Email: dakotah.s.martinez-1@ou.edu} and
Shadman Salam$^1$\footnote[4]{Email: shadman.salam@ou.edu} 
}\\ 
\vspace{1.2cm} \renewcommand{\thefootnote}{\arabic{footnote}}
{\it 
$^1$Homer L. Dodge Department of Physics and Astronomy,\\
University of Oklahoma, Norman, OK 73019, USA \\[3pt]
}
{\it 
$^2$Department of Physics,
University of Wisconsin, Madison, WI 53706 USA \\[3pt]
}

\end{center}

\vspace{0.5cm}
\begin{abstract}
\noindent
A vast array of (metastable) vacuum solutions arise from string
compactifications, each leading to different 4-d laws of physics.
The space of these solutions, known as the string landscape,
allows for an environmental solution to the cosmological constant problem.
We examine the possibility of an environmental solution to the
gauge hierarchy problem.
We argue that the landscape favors softly broken supersymmetric models over
particle physics models containing quadratic divergences,
such as the Standard Model. We present a scheme for computing relative
probabilities for supersymmetric models to emerge from the landscape.
The probabilities are related to the likelihood that the derived value of the
weak scale lies within the Agrawal et al. (ABDS) allowed window of values
leading to atoms as we know them.
This then favors natural SUSY models over unnatural (SUSY and other)
models via a computable probability measure.
\end{abstract}
\end{titlepage}

\section{Introduction}
\label{sec:intro}

Supersymmetry is a key ingredient in superstring theory constructs.
An advantage of compactification of 10-d string theory on a
Calabi-Yau manifold\cite{Candelas:1985en} is that it preserves some remnant
spacetime supersymmetry in the 4-d theory.
Likewise, compactification of 11-d M-theory on a manifold of
$G_2$ holonomy preserves some remnant spacetime supersymmetry\cite{Acharya:2001gy}.
Acharya\cite{Acharya:2019mcu} argues for the proposal that the landscape of all geometric,
stable, string/M theory compactifications to Minkowski spacetime at
leading order are supersymmetric.
Non-SUSY preserving compactifications would lead to bubble of nothing
instabilities and presumably lie within the swampland\cite{GarciaEtxebarria:2020xsr}.

Making contact with 4-d physics at the TeV scale (which is currently being explored at the CERN LHC), it is apparent that $N=1$ spacetime SUSY must be
broken.\footnote{For a recent review of the status of SUSY after LHC Run 2, see
\cite{Baer:2020kwz}.}
But the question is: broken at which scale? The gauge hierarchy problem
(GHP) suggests SUSY which is broken at or around the weak scale,
thus providing a ``natural'' solution to the GHP wherein all quadratically
divergent contributions to the Higgs boson mass cancel.
Weak scale supersymmetry is also supported experimentally via the measured
value of gauge couplings, whose values unify at a scale $m_{GUT}\simeq 2\times 10^{16}$ GeV\cite{Amaldi:1991cn} under renormalization group evolution\cite{Dimopoulos:1981yj} within the Minimal
Supersymmetric Standard Model\cite{Dimopoulos:1981zb} (MSSM) while they do not within the context of the Standard Model (SM). In addition, the measured value of the Higgs boson mass
falls directly within the narrow range of values allowed by the MSSM\cite{Slavich:2020zjv}.
Unfortunately, superpartners have so far failed to appear at LHC leading to
an apparent naturalness crisis\cite{Dine:2015xga}.

An apparent alternative to naturalness has emerged from the string
landscape\cite{Susskind:2003kw}.
Under flux compactifications\cite{Douglas:2006es}, an enormous number of
different compactification possibilities are available\cite{Bousso:2000xa,Ashok:2003gk},
each leading to different 4-d laws of physics.
Each of these possibilities can be accessed within
the context of an eternally-inflating multiverse\cite{Linde:2015edk}.
This scenario provides the proper setting to realize
Weinberg's anthropic solution to the cosmological constant
problem\cite{Weinberg:1987dv}: we find ourselves in a (pocket) universe
with a tiny cosmological constant $\Lambda_{cc}\sim 10^{-123}m_P^2$ because
if $\Lambda_{cc}$ was much larger, the universe would expand so fast that
structure (galaxies, stars, etc.) would not be able to form and life
as we know it would not arise. Such a solution to the CC problem is known as an
{\it environmental} (or {\it anthropic}) solution: environmental selection of
a tiny cosmological constant within the plenitude of pocket universes within
the greater multiverse can select a highly fine-tuned value for one (or more)
of our fundamental physical constants.
Such a solution may stand in apparent opposition to a natural solution
to the CC problem.

Can the GHP also be explained via environmental/anthropic reasoning rather
than naturalness? Maybe.
In the seminal papers by Agrawal, Barr, Donoghue and Seckel (ABDS)\cite{Agrawal:1997gf,Agrawal:1998xa},
the scenario of the Standard Model emerging from the multiverse via an
anthropic solution to the hierarchy problem is investigated. 
The authors consider the SM as the low energy effective field theory (LE-EFT),
but with a variable magnitude for the weak scale.
If the weak scale were a factor $\sim 2-5$
times larger than its actual value, then up-down quark mass differences
would increase, leading to nuclear instability: one enters a domain of the universe where only protons exist, with no complex nuclei.
For the weak scale reduced by a factor two from its measured value, then
protons become unstable and beta decay to neutrons:
there would be no Hydrogen, just neutron rich matter.
In terms of the Higgs vacuum expectation value $v$, one finds
$0.5\alt v/v_0\alt (2-5)$ (where $v_0$ is the Higgs vev in our universe).
This narrow range of values for the weak scale
has been dubbed the {\it ABDS window} in that values of $v$ outside this range
would not lead to a universe with life as we know it. The anthropic requirement
for $v$ to lie within the ABDS window could allow for a tuning of the weak scale
within the wider multiverse. It also selects out a narrow range of allowed values: namely $m_{weak}\simeq m_{W,Z,h}\sim 100$ GeV and can {\it explain}
the magnitude of the weak scale rather than just accommodate it.
The requirement for the magnitude to lie within the ABDS window is sometimes also referred to as the {\it atomic principle} in that it is required in order for any pocket universe to contain complex atoms which seem necessary for a
rich chemistry and for life as we know it.

Building upon the SM and ABDS, Arkani-Hamed and Dimopoulos\cite{Arkani-Hamed:2004ymt,Giudice:2004tc,Arkani-Hamed:2004zhs}
proposed a model known as Split Supersymmetry wherein the natural
SUSY solution to the GHP is eschewed in exchange for an
environmental solution.
This then allows the possibility of a highly fine-tuned supersymmetric model.
The authors then investigate the consequences of scalar masses $\tilde{m}$
far beyond the naturalness limit, taking $\tilde{m}$ as high as $\sim 10^9$ GeV.
SUSY fermions, higgsinos and gauginos, may be protected by a chiral or $R$-symmetry and may still live around the EW scale. This set-up maintains the successful gauge coupling unification and WIMP dark matter of SUSY models,
but enlists the vast number of landscape solutions to effectively
tune the weak scale to lie within the ABDS window as required by the
atomic principle. The advantage of very heavy scalars (especially first/second generation matter scalars), as noted much earlier
by Dine {\it et al.}\cite{Dine:1990jd} and others\cite{Cohen:1996vb,Bagger:1999sy} is that they provide a decoupling solution to the SUSY flavor and CP
problems and may also suppress proton decay.
In addition, under gravity mediation wherein scalars get mass of order the
gravitino mass, this provides a solution to the cosmological gravitino and
moduli problems.\footnote{For a recent overview of the cosmological
  moduli problem, see {\it e.g.} \cite{Bae:2022okh}.}

Thus, Split SUSY and a variety of successor models\cite{Hall:2011jd,Arvanitaki:2012ps} have been considered
as legitimate expressions of what sort of SUSY models are
expected to emerge from the string landscape.
In the literature, it is sometimes claimed that a rather heavy Higgs
mass and no sign of SUSY scalars at LHC might be construed as evidence for
finetuning within the multiverse as opposed to a natural solution to the GHP,
wherein there is no finetuning.
Split SUSY, and the other high-scale SUSY models considered here,
are motivated by the expectation that the soft SUSY breaking terms are
statistically favored to occur at large as opposed to small values on the
landscape via a power law relation $P(m_{soft})\sim m_{soft}^{2n_F+n_D-1}$
which obtains if the complex-valued SUSY breaking $F$-term fields and
real-valued SUSY breaking $D$-term fields are distributed
uniformly on the landscape\cite{Douglas:2004qg,Susskind:2004uv,Arkani-Hamed:2005zuc}.
(Here, $n_F$ is the number of hidden sector
$F$-term fields and $n_D$ is the number of hidden sector $D$-term fields
contributing to the overall SUSY breaking scale.)
This landscape draw to large soft terms must be
balanced by the anthropic/environmental condition that the derived value
of the weak scale in each pocket universe lies within the ABDS window
of allowed values\cite{Baer:2016lpj,Baer:2017uvn}.

In this paper we survey a variety of
finetuned models (both the SM and SUSY), and compare these to natural
SUSY models, all within the context of the string landscape.
What we find is somewhat at odds with the literature: natural
SUSY models are more likely to emerge from the string landscape than
finetuned models. We advance a particular probability measure $P_\mu$
which quantifies these probabilities.
By taking ratios, we are able to evaluate the relative probabilities for
different models to emerge from the landscape.

In our deliberations, weak scale naturalness plays a key role, and we must
define what we mean by naturalness. We adopt the definition of so-called
{\it practical naturalness}\cite{Baer:2015rja}: an observable ${\cal O}$ is natural provided
that all {\it independent} contributions to ${\cal O}$ are comparable to
or less than ${\cal O}$. For the case of the SM, where the Higgs potential
is given by
\be
V=-\mu_{SM}^2\phi^\dagger\phi +\lambda (\phi^\dagger\phi)^2
\ee
a vacuum expectation value $v=\sqrt{\mu_{SM}^2/\lambda}$ develops and
the tree-level Higgs boson mass is given by $m_h^2=2\mu_{SM}^2$.
The loop-corrected Higgs mass is quadratically divergent up to some
cutoff scale $\Lambda_{SM}$ where 
\be
m_h^2=2\mu_{SM}^2+\delta m_h^2
\label{eq:mhs}
\ee
where at one loop
\be
\delta m_h^2\simeq\frac{3}{4\pi^2}\left(-\sum_i\lambda_i^2+\frac{g^2}{4}+\frac{g^2}{8\cos^2\theta_W}+\lambda\right)\Lambda_{SM}^2
\label{eq:deltamhs}
\ee
where the $\lambda_i$ are Yukawa couplings for the $i$th fermion, $g$ is the $SU(2)_L$ gauge coupling and $\lambda$ is the Higgs quartic coupling\cite{Giudice:2008bi}.
Requiring practical naturalness then leads to $\Lambda_{SM}\alt 1$ TeV whilst
finetuning is required for much higher values of $\Lambda_{SM}\gg 1$ TeV.

In SUSY models with the MSSM as the LE-EFT, then the weak scale is actually
predicted in terms of the weak scale soft SUSY breaking terms and
superpotential $\mu$ parameter. Minimization of the Higgs potential in the
MSSM leads to
\be
m_Z^2/2=\frac{m_{H_d}^2+\Sigma_d^d-(m_{H_u}^2+\Sigma_u^u)\tan^2\beta}{\tan^2\beta-1}-\mu^2\simeq -m_{H_u}^2-\mu^2-\Sigma_u^u(\tst_{1,2}) 
\label{eq:mzs}
\ee
where the right-hand-side approximation is obtained for moderate-to-large
$\tan\beta \agt 5$.
Here, $m_{H_u}^2$ and $m_{H_d}^2$ are the Higgs soft breaking masses and $\tan\beta$
is the usual ratio of Higgs field vacuum expectation values $v_u/v_d$.
The $\Sigma$ terms contain over 40 loop corrections (explicit formulae for
the $\Sigma$ terms may be found in Ref's \cite{Baer:2012cf}
and \cite{Baer:2021tta} and leading two-loop terms may be found in
Ref. \cite{Dedes:2002dy}). Requiring the MSSM weak scale as given by the measured value of $m_Z$ to be natural then requires $|\mu|\alt 350$ GeV while
$m_{H_u}^2$ is driven radiatively to small negative values at the weak scale (electroweak
symmetry is barely broken). Also, the leading loop corrections $\Sigma_u^u(\tst_{1,2})$ are minimized for TeV scale top squarks with large, negative
$A_t$ trilinear soft terms\cite{Baer:2012up} which also give rise to nearly maximal stop
mixing and large values of $m_h\sim 125$ GeV. The finetuning measure\cite{Baer:2012up}
\be
\Delta_{EW}\equiv |maximal\ term\ on\ RHS\ of\ Eq.\ \ref{eq:mzs}|/(m_Z^2/2)
\label{eq:dew}
\ee
is then a measure of practical naturalness in the MSSM, where for natural
models usually $\Delta_{EW}\alt 30$ is required.
Notice that $\Delta_{EW}$ is closely related to the ABDS anthropic window in
that requiring $\Delta_{EW}\alt 30$ then requires all independent
contributions to the MSSM weak scale to be within the ABDS window.

\section{A survey of some natural and unnatural SUSY models}
\label{sec:models}

\subsection{CMSSM}

For a long time, the mSUGRA\cite{Chamseddine:2000nk} or
CMSSM\cite{Kane:1993td} model served as a sort of paradigm model
for SUSY phenomenology.
This model posits gravity-mediated SUSY breaking which induces a
common scalar mass $m_0$, a common gaugino mass $m_{1/2}$ and a common
trilinear soft term $A_0$ all prescribed at the GUT
scale $m_G\simeq 2\times 10^{16}$ GeV.
The weak scale soft terms are determined by RGE running to the weak scale,
where electroweak symmetry is radiatively broken via a large top quark
Yukawa coupling.
The $\mu$ term is tuned via Eq. \ref{eq:mzs} to give the measured value of $m_Z$.
In pre-LHC days, it was possible within the CMSSM model to gain accord with
naturalness (low $\Delta_{EW}$) and with an acceptable thermal relic abundance
of the LSP.
After LHC Run 2-- while respecting the LHC measured Higgs mass and also
LHC sparticle search limits-- natural CMSSM spectra are no longer possible\cite{Baer:2012mv,Baer:2014ica,Baer:2019cae}.

For illustrative purposes, we compute the mSUGRA/CMSSM spectra using the
Isasugra spectrum generator\cite{Paige:2003mg,Baer:1994nc} for a mSUGRA/CMSSM benchmark
point with ($m_0,\ m_{1/2},\ A_0,\ \tan\beta =5000\ {\rm GeV},1200\ {\rm GeV},\ -8000\ {\rm GeV},10$) which yields a gluino mass $m_{\tg}=2.8$ TeV (well above current LHC bounds) with $m_h=124.3$ GeV and with $\Delta_{EW}=2641$ (highly EW finetuned). The thermal bino LSP abundance is $\Omega_\chi h^2\simeq 249$ so
non-thermal processes would need to be invoked to bring the relic density
into alignment with the measured dark matter abundance\cite{Baer:2014eja}.

\subsection{PeV SUSY}

PeV-scale supersymmetry\cite{Wells:2003tf,Wells:2004di} is motivated by the possibility of SUSY breaking
via ``charged'' SUSY breaking fields $S$. For charged SUSY breaking,
scalar partners gain mass via K\"ahler potential terms
$K\ni\frac{S^\dagger S}{m_P^2}Q^\dagger Q$ where the $Q$ are visible sector fields and $S$ are hidden sector fields which carry some charge, perhaps $R$ charge.
Thus, scalar fields gain a mass $m_Q^2\sim F_S^\dagger F_S/m_P^2\sim m_{3/2}^2$
whilst
gaugino masses, which ordinarily gain mass via the gauge kinetic function
$f\ni kS$ are forbidden. Hence, the leading contribution to gaugino masses
(and also $A$-terms) are the loop-suppressed anomaly-mediated contributions
$m_\lambda=\frac{\beta(g_\lambda )}{g_\lambda}m_{3/2}$ and we expect
$M_1\simeq m_{3/2}/120$, $M_2\simeq m_{3/2}/360$ and $M_3\simeq m_{3/2}/40$.
The wino is then the LSP and can make up the dark matter. Thermally produced
relic winos can make up all the missing dark matter for $m_{wino}\sim 3$ TeV.
Then, with a 3 TeV wino, one expects scalar masses
$\tilde{m}\sim 1000$ TeV, {\it i.e.} close to the PeV scale (1 PeV=1000 TeV).
The PeV scale scalar masses provide a decoupling solution to the SUSY
flavor and CP problems\cite{Dine:1990jd}.
The $\mu$ parameter may range anywhere between $m_{wino}$ and $\tilde{m}$.
The resultant light Higgs mass is expected in the range
$125\ {\rm GeV} <m_h<155\ {\rm GeV}$\cite{Giudice:2004tc}.

\subsection{Split SUSY} 

In split SUSY\cite{Arkani-Hamed:2004ymt,Giudice:2004tc,Arkani-Hamed:2004zhs},
the motivation is that the string landscape may provide a
selection mechanism for the finetuning of the electroweak scale in that
the weak scale must lie within the ABDS window in order to have a universe
with complex atoms as we know them, which seem necessary for life.
However, SUSY may still be needed for consistency with string theory,
but the SUSY breaking scale may now be far higher than that which is
usually required by naturalness.
One may then allow masses of squarks and sleptons
(which occur in multiplets of $SU(5)$) to be as high as $m_\phi\sim 10^9$ GeV
while fermion masses, which are protected by chiral symmetry, can lie
near the weak scale. This model then preserves the SUSY success stories of
gauge coupling unification and WIMP dark matter while appealing to vacuum
selection from the string landscape to ``tune'' the EW scale to its
value as required by the atomic principle.
Thus, in split SUSY, one expects both gauginos and higgsinos around the
weak scale whilst squarks and sleptons decouple at some intermediate scale
({\it e.g.} $10^9$ GeV). Such a split hierarchy of masses can arise
from $D$-term SUSY breaking which maintains an approximate,
accidental $R$-symmetry\cite{Arkani-Hamed:2004zhs}.
The very high scalar mass scale $\tilde{m}$ provides a decoupling solution
to the SUSY flavor and CP problems and also alleviates the cosmological
gravitino and moduli problems by making these particles sufficiently
heavy and thus shortlived in the early universe.
The striking signature of split SUSY models is long lived gluinos
which may decay with displaced vertices or even outside of the collider
detector.
For scalar masses as high as $\sim 10^9$ GeV, then the lightest Higgs scalar is
expected to have mass $m_h\sim 130-145$ GeV\cite{Giudice:2011cg}.

\subsection{High-scale SUSY}

In high-scale SUSY (HS-SUSY)\cite{Barger:2005qy,Barger:2007qb,Ellis:2017erg},
it is assumed that the underlying 4-d theory is indeed SUSY,
but with a much higher SUSY breaking scale than that which is usually
assumed to solve the gauge hierarchy problem.
Thus, in HS-SUSY, the superpartners
are typically clustered at some very high mass scale $\tilde{m}\sim 10-10^{13}$ TeV. In HS-SUSY, the SM is the LE-EFT and only the light Higgs particle is
expected to be produced at LHC. Indeed, by requiring the model to yield the
measured Higgs mass $m_h\sim 125$ GeV, then $\tilde{m}\sim 10^1-10^7$ TeV\cite{Giudice:2011cg,Bagnaschi:2014rsa}.

\subsection{Mini-Split}

Mini-Split\cite{Arvanitaki:2012ps} SUSY is a version of split SUSY
wherein the scalar mass $\tilde{m}$ is lowered to the
$\sim 10^{2-4}$ TeV range in order to accommodate the measured Higgs mass
$m_h\simeq 125$ GeV while gauginos remain near the TeV scale. Several scenarios
are envisaged in \cite{Arvanitaki:2012ps} including non-sequestered AMSB and $U(1)^\prime$
mediation. These scenarios include a small $A$ parameter while $\mu$ may be
either at the gaugino scale (light) or at the scalar scale (heavy).

\subsection{Simply unnatural SUSY}

In simply unnatural SUSY\cite{Arkani-Hamed:2012fhg} (SUN-SUSY),
the scalar mass scale $\tilde{m}$ is determined
by the measured value of the Higgs mass $m_h\simeq 125$ GeV to be
$\tilde{m}\sim 10^2-10^3$ TeV where the trilinear soft terms $A_i$ are assumed to be
tiny (little mixing in the stop sector). The SUSY $\mu$ term is also expected
to be $\mu\sim\tilde{m}$ while the gaugino masses, which require an
$R$-symmetry breaking to gain mass, are expected to be at the TeV scale.
Minimally, the gaugino masses are expected to obtain the AMSB form, but the
presence of heavy vector-like states could alter those relations leading
to a more compressed gaugino spectrum.
Typically, the wino is expected to be the LSP,
and the relic abundance may be produced either thermally or non-thermally
due to late-decaying TeV-scale moduli fields.

\subsection{Spread SUSY} In Ref.~\cite{Hall:2011jd},
it is emphasized that there may exist
a forbidden region on the scale of SUSY breaking $\tilde{m}$ such that if
$\tilde{m}\agt {\cal O}(1)$ TeV, then LSP dark matter will be overproduced
which can violate the anthropic bounds which disfavor/forbid DM overproduction
in that the baryon-to-DM ratio may be insufficient for baryonic structure formation in the universe\cite{Tegmark:2005dy}.
This forbidden region should persist up to
$\tilde{m}\sim T_R$ where $T_R$ is the reheat temperature of the universe at the end of inflation. Higher values of $\tilde{m}>T_R$ are allowed in that
SUSY particles wouldn't be produced during the reheat process.
Taking $\tilde{m}>T_R$ then leads to a very heavy SUSY spectrum
(High Scale SUSY) whilst taking $\tilde{m}\sim 1$ TeV leads to Spread SUSY
in the case of SUSY breaking via ``charged'' hidden sector fields (where
scalars gain mass $\tilde{m}$ but gauginos and $A$ terms do not) or via
uncharged hidden sector fields (which leads to all sparticles at
$\tilde{m}\sim 10$ TeV, dubbed the ``environmental MSSM''). The spread SUSY
spectrum divides into two possibilities: 1. scalar masses $\tilde{m}\sim 10^5$
TeV with gauginos at $10^2$ TeV and higgsinos at $\sim 1$ TeV and 2.
scalars around $10^3$ TeV with higgsinos and gravitinos $\sim 10^2$ TeV
and gauginos $\sim 1$ TeV. Thus, the spread SUSY models typically have SUSY mass
spectra spread across {\it three} mass scales.

\subsection{G$_2$MSSM} The G$_2$MSSM labels the sort of SUSY spectra expected
to emerge from 11-dimensional $M$-theory compactified on a manifold of
$G_2$ holonomy\cite{Acharya:2006ia,Acharya:2008zi} which preserves $N=1$ SUSY
in the low energy 4-d effective field theory (LE-EFT).
The LE-EFT then consists of the usual MSSM fields plus an assortment of moduli
fields which are string remnants from the compactification.
Scalar masses $\tilde{m}$ 
and the lightest modulus field are expected to gain masses of order the
gravitino mass $m_{3/2}$ and in order to solve the cosmological
moduli/gravitino problems then $\tilde{m}\sim 30-100$ TeV.
Gaugino masses are suppressed relative to scalars by a factor
$\log (m_P/m_{3/2})\sim 30$ so gauginos (and higgsinos)
are expected at the 1-3 TeV range
and may have comparable moduli/anomaly-mediated contributions.
The LSP may be bino or wino-like but the relic density is seriously affected
by non-thermal production via the late-decaying lightest modulus
field\cite{Acharya:2008bk}.
In later renditions, the possibility of a hidden sector LSP is
entertained\cite{Acharya:2016fge}.

\subsection{Radiatively-driven/Stringy natural SUSY}

In radiatively-driven natural SUSY (RNS)\cite{Baer:2012up,Baer:2012cf},
large high scale soft terms can be radiatively driven to small weak
scale values. 
Then all weak scale contributions to the weak scale are of order the weak
scale.
This corresponds to
$\Delta_{EW}\alt 30$.\footnote{A computer code {\bf DEW4SLHA} is
  available which computes the value of $\Delta_{EW}$ for any MSSM model
  listed in SUSY Les Houches Accord format\cite{Baer:2021tta}.}
The RNS models can be generated from NUHM2 or NUHM3 models\cite{Baer:2012up,Baer:2012cf}, from
generalized mirage mediation\cite{Baer:2016hfa} and from natural anomaly-mediation\cite{Baer:2018hwa}.
As an example, we take a simple NUHM2 model with first/second/third
generation GUT scale scalar masses $m_0(1,2)=m_0(3)=4.5$ TeV, 
$m_{1/2}=1$ TeV, $A_0=-7.2$ TeV, $\tan\beta =10$ with $\mu =200$ GeV and $m_A=2$ TeV. The model has $m_{\tg}\sim 2.4$ TeV (LHC safe) with $\Delta_{EW}=12.8$
and $m_h=124.3$ GeV.
The higgsino-like LSP is $m_\chi=195.3$ GeV with $\Omega_\chi h^2=0.011$
(so room for additional axion dark matter).

While RNS models are typically slightly more natural for lower $m_0$ and $m_{1/2}$ values, we expect from the string landscape, under spontaneous
SUSY breaking via a single $F$-term field distributed uniformly as a
complex number throughout the landscape, a {\it linear} statistical
draw to large soft terms\cite{Baer:2016lpj}.
For more SUSY breaking fields, the statistical
draw goes as $f_{SUSY}\sim m_{soft}^{2n_F+n_D-1}$ where $n_F$ is the number of hidden sector $F$ breaking fields and $n_D$ is the number of hidden sector $D$-breaking fields (the latter distributed as real numbers)\cite{Douglas:2004qg,Susskind:2004uv,Arkani-Hamed:2005zuc}.\footnote{In \cite{Broeckel:2020fdz}, it is found that a linear
$n=1$ soft term draw is obtained for KKLT\cite{Kachru:2003aw} moduli-stabilization models.}
Convolution of the statistical
draw to large soft terms with the anthropic requirement that the
derived weak scale lies within the ABDS window then leads to a
probability distribution for $m_h$ that rises to a peak around
$m_h\sim 125$ GeV\cite{Baer:2017uvn} (in part because $A_0$ is also drawn to
large (negative) values giving maximal stop mixing\cite{Baer:2011ab})
with sparticles typically beyond LHC reach.
In this rendition,
naturalness is replaced by what Douglas calls {\it stringy naturalness}\cite{Douglas:2004zg},
where a mode is more stringy natural if more landscape vacua lead to
such a result. In stringy natural SUSY, a 3 TeV gluino is more (stringy)
natural than a 300 GeV gluino\cite{Baer:2019cae}.
The RNS benchmark given above is thus highly stringy natural.
Thus, under stringy naturalness, RNS models with LHC-compatible sparticle
masses most commonly emerge from the landscape\cite{Baer:2022wxe}.

\section{A scheme for computing relative probabilities from the landscape}

The central question we wish to address is: how likely are various
SUSY models (and the SM) to arise from the landscape?
To answer this, we will restrict ourselves to string vacua containing the
MSSM as the low energy EFT, and where SUSY breaking is mediated by gravity,
{\it i.e.} spontaneous SUSY breaking in a $N=1$ supergravity framework.\cite{Cremmer:1982en}
In such a SUGRA framework, scalar masses are generically
{\it non-universal}\cite{Soni:1983rm,Kaplunovsky:1993rd,Brignole:1993dj,Baer:2020vad}
unless protected by some symmetry: {\it e.g.} the matter scalars of each
generation fill the 16-dimensional spinor rep of $SO(10)$ so one might expect
these to have a common mass $m_0(i)$, $i=1-3$ a generation index.\footnote{We regard the AMSB soft terms as included in the gravity-mediated soft terms.}
Since the Higgs scalars come in split multiplets, there is no reason to
expect $m_0(i)=m_{H_{u,d}}$ and thus we expect the LE-EFT to be a
non-universal Higgs model (NUHM).
This framework accommodates all of the
high-scale and natural SUSY models under consideration here.\footnote{
  For instance, in this framework, there is no known reason to favor
  the CMSSM model over any of the NUHM models.} 
While an absolute probability for any particular LE-EFT (including those
not within the realm of the MSSM) is not possible to calculate
(at least at this time), we can make estimates of {\it relative} probabilities
amongst gravity-mediated MSSM models based on certain reasonable
assumptions.

\begin{figure}[!htbp]
\begin{center}
\includegraphics[height=0.6\textheight]{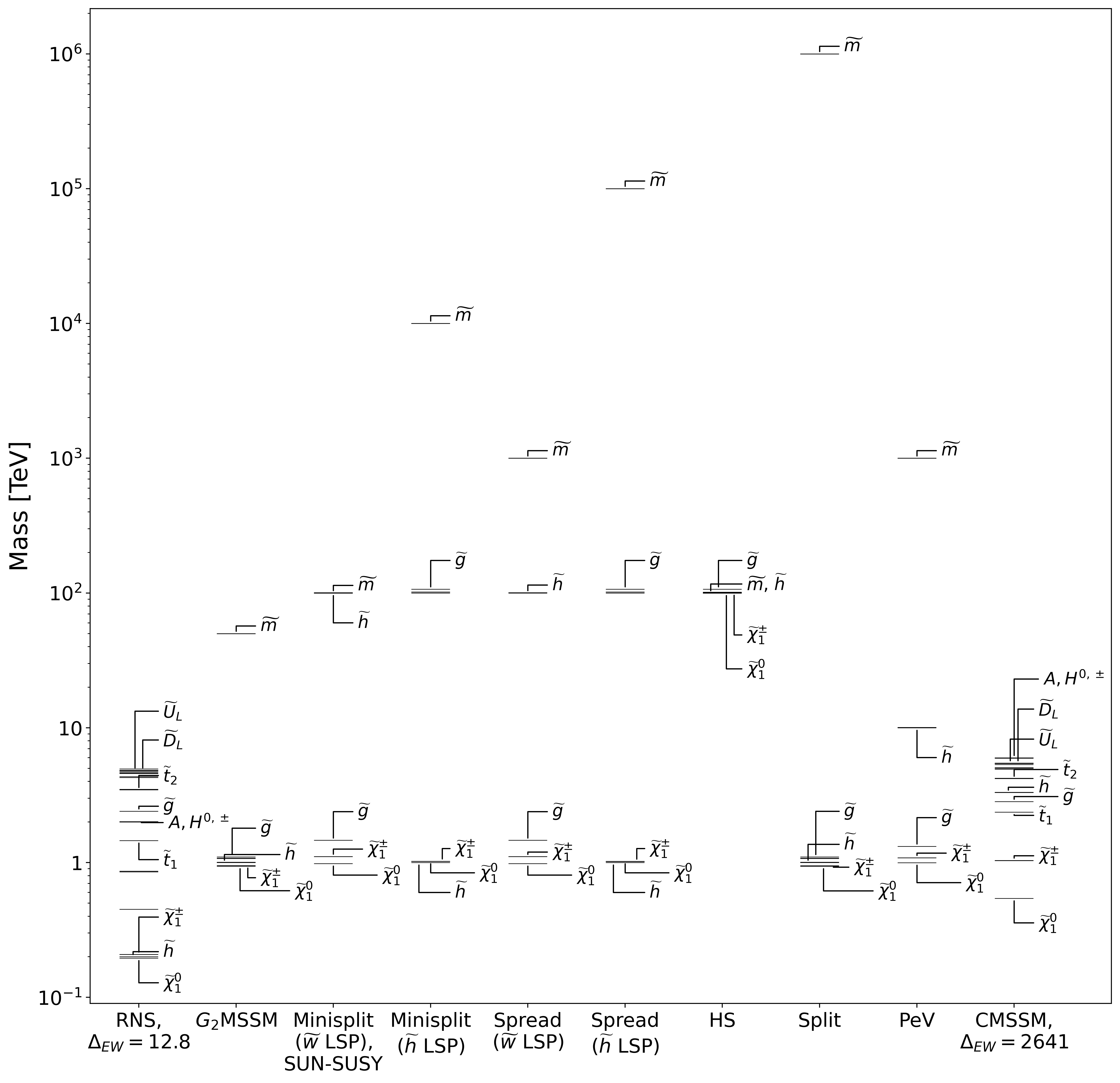}
\caption{Mass spectra from various unnatural and natural SUSY models
  as depicted in Table \ref{tab:models}.
  \label{fig:spectra}}
\end{center}
\end{figure}

In Table \ref{tab:models}, we list a variety of supersymmetric models,
along with the SM, and the proposed range for various first/second $m_0(1,2)$
and third generation $m_0(3)$ scalar masses, along with the expected
range for gaugino and higgsino masses and the range of the light Higgs mass.
In the last column we list the relative probability measure $P_\mu$ to be
explained below. For the two SUSY models CMSSM and RNS, we have approximate supersymmetry extending all the way down to the weak scale.
For the remainder of SUSY models, which include rather high mass scales
$\tilde{m}$, we assume the heavy SUSY states are integrated out at
scale $Q\simeq \tilde{m}$ which then destroys softly broken SUSY below the
$\tilde{m}$ scale, so that quadratic divergences arise which are proportional
to $\Lambda=\tilde{m}$ as in Eq. \ref{eq:deltamhs}.
A pictorial comparison of the spectra from the various models is given in
Fig. \ref{fig:spectra}.
\begin{table}\centering
\begin{tabular}{lcccccc}
\hline
model & $\tilde{m}(1,2)$ & $\tilde{m}(3)$ & gauginos & higgsinos & $m_h$ & $P_\mu$ \\
\hline
SM & - & - & - & -& - & $7\cdot10^{-27}$ \\
CMSSM ($\Delta_{EW}=2641$) & $\sim 1$ & $\sim 1$ & $\sim 1$ & $\sim 1$ & $0.1-0.13$
& $5\cdot 10^{-3}$ \\
PeV SUSY & $\sim 10^3$ & $\sim 10^3$ & $\sim 1$ & $1-10^3$ &
$0.125-0.155$ & $5\cdot 10^{-6}$ \\
Split SUSY & $\sim 10^6$ & $\sim 10^6$ & $\sim 1$ & $\sim 1$ & $0.13-0.155$
& $7\cdot 10^{-12}$ \\
HS-SUSY & $\agt 10^2$ & $\agt 10^2$ & $\agt 10^2$ & $\agt 10^2$ & $0.125-0.16$
& $6\cdot 10^{-4}$ \\
Spread ($\tilde{h}$LSP) & $10^{5}$  & $10^5$ & $10^2$ & $\sim 1$ & $0.125-0.15$ & $9\cdot 10^{-10}$ \\
Spread ($\tilde{w}$LSP) & $10^{3}$ & $10^{3}$ & $\sim 1$ & $\sim 10^2$ & $0.125-0.14$  & $5\cdot 10^{-6}$ \\
Mini-Split ($\tilde{h}$LSP)& $\sim 10^4$ & $\sim 10^4$ & $\sim 10^2$ & $\sim 1$  & $0.125-0.14$ & $8\cdot10^{-8}$ \\
Mini-Split ($\tilde{w}$LSP)& $\sim 10^2$ & $\sim 10^2$ & $\sim 1$ & $\sim 10^2$ & $0.11-0.13$ & $4\cdot 10^{-4}$ \\
SUN-SUSY  & $\sim 10^2$ & $\sim 10^2$ & $\sim 1$ & $\sim 10^2$  & $0.125$
& $4\cdot 10^{-4}$ \\
G$_2$MSSM  & $30-100$ & $30-100$ & $\sim 1$  & $\sim 1$  & $0.11-0.13$
& $2\cdot 10^{-3}$ \\
RNS/landscape & $5-40$  & $0.5-3$ & $\sim 1$ & $0.1-0.35$ & $0.123-0.126$
& $1.4$ \\
\hline
\end{tabular}
\caption{A survey of some unnatural and natural SUSY models
  along with general expectations for sparticle and Higgs
  mass spectra in TeV units.
  We also show relative probability measure $P_\mu$ for the model to emerge
  from the landscape.
  For RNS, we take $\mu_{min}=10$ GeV.
}
\label{tab:models}
\end{table}

For the two SUSY models RNS and CMSSM, the dominant contribution to the weak
scale can be extracted from the value of $\Delta_{EW}$.
Then the pocket universe value of $m_Z^{PU}$ can be computed using
Eq.~\ref{eq:mzs} as
\be
\frac{(m_Z^{PU})^2}{2}=\frac{(m_Z^{OU}\sqrt{\Delta_{EW}})^2}{2}-\mu_{PU}^2
\ee
(assuming the dominant contribution dominates all other contributions
to $(m_Z^{PU})^2$, which is usually the case.) Here, $m_Z^{OU}=91.2$ GeV,
the value of $m_Z$ in our universe (OU). In most SUSY spectrum calculations,
the value of the $\mu$ parameter is finetuned to ensure that $m_Z$ gains
its measured value in our universe.
However, in the multiverse, each pocket universe containing the MSSM as the
LE-EFT will have a different value of $\mu_{PU}$ which will in general
lead to a value for the weak scale which is very different from the one
in our universe: $m_Z^{PU}\ne m_Z^{OU}$. In fact, frequently $m_Z^{PU}$ may differ
from $m_Z^{OU}$ by many orders of magnitude.
If it does, then one will have a pocket universe with $m_{weak}$ outside the
anthropic ABDS window, thus violating the atomic principle.

What is the likely distribution of SUSY $\mu_{PU}$ parameters in the multiverse?
Here, we assume the $\mu$ parameter arises in the superpotential
as in the Kim-Nilles (KN) solution to the SUSY $\mu$ problem\cite{Kim:1983dt},\footnote{
  For a recent review of twenty solutions to the SUSY $\mu$ problem, see
  Ref. \cite{Bae:2019dgg}.}
where we expect $W\ni \lambda_\mu S^2H_uH_d/m_P$. The PQ charged field $S$
acquires a vev of order $f_a\sim 10^{11}$ GeV under PQ breaking so that a
$\mu$ parameter arises:
\be
\mu (KN)\sim \lambda_\mu f_a^2/m_P\sim m_{weak} .
\ee
Thus, the KN $\mu$ parameter has the form of a (Planck-suppressed)
Yukawa coupling, in accord with the other Yukawa couplings which occur
in the superpotential.
But the question is: what sort of distribution for $\mu$ would we
expect on the landscape? For fixed $\lambda_\mu$, this has been computed
in a particular well-motivated KN solution based on an anomaly-free
discrete $R$-symmetry ${\bf Z}_{24}^R$\cite{Baer:2021vrk}.
However, for non-fixed $\lambda_{\mu}$, this may well be different. In fact,
Donoghue, Dutta and Ross\cite{Donoghue:2005cf} make a convincing case
that Yukawa couplings are distributed uniformly across the decades of
possible values, which appears to match well with the measured
fermion mass values. We will adopt the Donoghue {\it et al.} ansatz for the
KN $\mu$ parameter as well: that no particular scale for the $\mu_{PU}$
value is favored over any other from the string landscape.
This seems reasonable in that the only scale inherent in string theory
is the string scale, and all other scales likely arise dynamically:
{\it i.e.} there is no preferred scale for $\mu_{PU}$.
This corresponds to a landscape
distribution $f_\mu\sim 1/\mu$ so that the integrated distribution is indeed
scale invariant.

\begin{figure}[!htbp]
\begin{center}
\includegraphics[height=0.5\textheight]{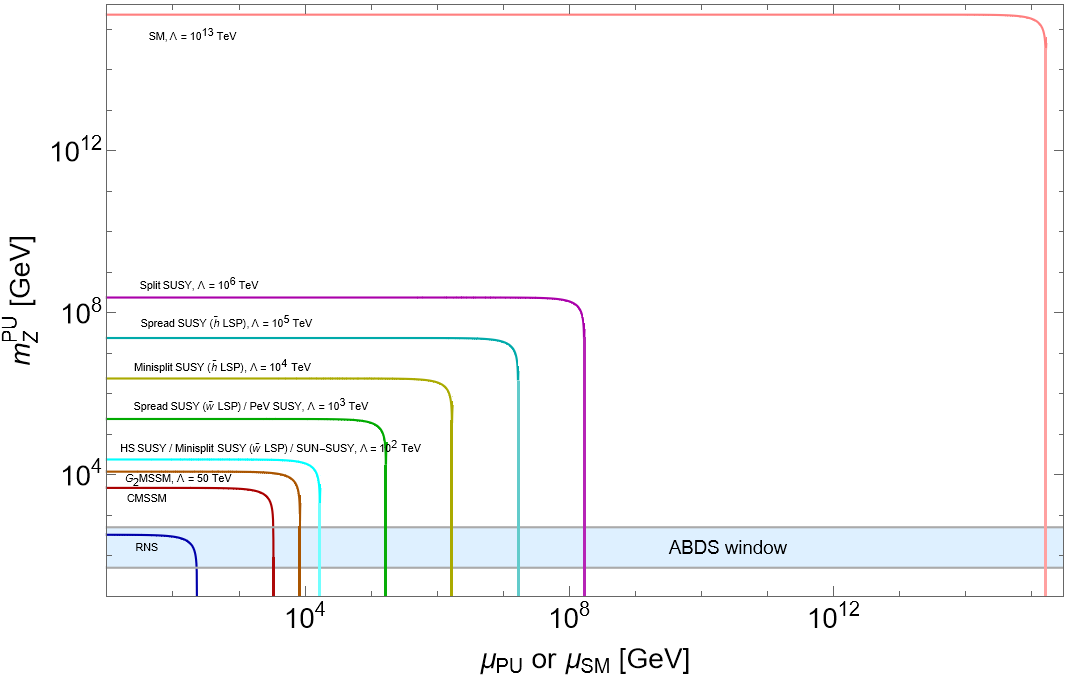}
\caption{Values of $m_Z^{PU}$ vs. $\mu_{PU}$ or $\mu_{SM}$ for
  various natural (RNS) and unnatural SUSY models and the SM.
  The ABDS window extends here from $m_Z^{PU}\sim 50-500$ GeV.
\label{fig:mzPU}}
\end{center}
\end{figure}

In Fig. \ref{fig:mzPU}, we show on the $x$-axis over 15 decades of
possible values for $\mu_{PU}$. For the RNS model, where the maximal
contribution to the RHS of Eq. \ref{eq:mzs} is bounded to lie within a factor a
few of our measured value of the weak scale, then there is a
substantial range of $\mu_{PU}$ values leading to $m_Z^{PU}$ lying within the
(blue-shaded) ABDS window. We will take (quite arbitrarily) the lower limit
of $\mu_{PU}$ to be $\sim 10$ GeV.
Values of $\mu_{PU}(min)$ higher or lower by an order of magnitude from
this value lead to differences in $P_\mu$ of a factor $\sim 2$
which is inconsequential for our purposes. 
The probability for a random value of $\mu_{PU}$ to give rise to
$m_Z^{PU}$ within the ABDS window is then
\be
P_\mu \equiv \log_{10}(\mu_{PU}(max)/\mu_{PU}(min))
\ee
\label{eq:Pmu}
   {\it i.e.} the length of the interval of logarithmically distributed
   $\mu_{PU}$ values.
Using this interval, we find $P_\mu(RNS)\sim 1.4$. 

For the CMSSM benchmark model with $\Delta_{EW}=2641$, then the maximal
contribution to the RHS of Eq. \ref{eq:mzs} is well beyond the ABDS window.
Thus, a finely-tuned value of $\mu_{PU}$ will be needed in order for
$m_Z^{PU}$ to lie within the ABDS window, in accord with the atomic principle.
One will have to live in the nearly vertical portion of the red CMSSM curve,
for which the interval length is $P_\mu (CMSSM)\sim 0.005$.
While the absolute values of $P_\mu$ don't have a particular meaning
(we don't know the overall normalization),
the ratios of probabilities do.
In this case, we would expect the RNS model to be
$P_\mu(RNS)/P_\mu(CMSSM)\sim 260$
times more probable on the landscape than the CMSSM benchmark model.
In this case, the ``natural'' value for the weak scale in the case of the CMSSM
benchmark model would be $m_{weak}\sim m_Z\sqrt{\Delta_{EW}}\sim 5$ TeV.

We can also calculate a value of $P_\mu$ for the SM, assuming the SM is valid
all the way up to a scale $Q\sim m_{GUT}$ as is assumed in estimates of the
SM vacuum stability\cite{Degrassi:2012ry}. Here, we will also assume that
$\mu_{SM}$ has a scale invariant distribution so that the $x$-axis of
Fig. \ref{fig:mzPU} pertains to $\mu_{SM}$ of Eq. \ref{eq:mhs} as well as to
$\mu_{PU}$. Taking the value of $m_Z^{PU}\sim m_h^{PU}$,
we can use Eq. \ref{eq:mhs} to plot the value of the weak scale in the SM.
The plot is shown in Fig. \ref{fig:mzPU} as the SM curve.
Here, we see a value of $\mu_{SM}\sim 10^{15}$ GeV
is needed for $m_Z^{PU}(SM)$ to lie within the ABDS window while the natural
value of $m_Z^{PU}(SM)$ is $\sim 10^{15}$ GeV. This shows the extreme finetuning
needed by the SM in order to ensure the weak scale lies within the ABDS window.
We can compute $P_\mu (SM)$ and find it to be $\sim 7\cdot 10^{-27}$, that is
the RNS model about $10^{26}$ times more likely than the SM to emerge
from the landscape.

We can now also compute $P_\mu$ values for the various high-scale SUSY models
listed in Table \ref{tab:models}.
The key point here is that quadratic divergences still cancel out at energy
scales $Q>\tilde{m}$. But once $Q$ drops below $\tilde{m}$, then we must
integrate out the heavy sparticles in the LE-EFT and the quadratic divergences
no longer cancel. Then we may use the uncanceled terms in
Eq. \ref{eq:deltamhs} to compute corrections to the Higgs mass, again with
$m_Z^{PU}\simeq m_h$. For most of these models, we take
$\Lambda\sim \tilde{m}=m_0(3)$ to compute the curves of $m_Z^{PU}$ vs.
$\mu_{SM}$, where now the Higgs potential is that of the SM for
$Q<\tilde{m}$.\footnote{For the SM parameter values entering Eq. \ref{eq:deltamhs} in the case of high scale SUSY models with scale boundary $\tilde{m}$, we
  use FlexibleSUSY and FlexibleEFTHiggs to extract the appropriate values
  \cite{Athron:2014yba,Athron:2016fuq,Athron:2017fvs}.}

The various curves are shown in Fig. \ref{fig:mzPU}
for the assorted high scale SUSY models of Table \ref{tab:models}.
We can then extract the values of $P_\mu$ for each case.
As an example, Split SUSY with $m_0(3)\sim 10^6$ TeV gives
$P_\mu\sim 7\cdot 10^{-12}$ so that RNS is $\sim 10^{12}$ times more likely than
Split SUSY to emerge from the landscape.
Lest one be dismayed by the low relative probability for Split SUSY to
emerge from the landscape, it is worth noting that the Split SUSY benchmark is
$\sim 10^{15}$ times more likely to emerge from the landscape than the SM
(when the SM is valid up to $Q=m_{GUT}$).
Scaling $\tilde{m}$ to lower values
in order to accommodate the measured value of $m_h$ as in mini-split helps
matters somewhat: in this case, mini-split with a wino LSP and
$\tilde{m}\sim 10^2$ TeV has $P_\mu\sim 4\cdot 10^{-4}$,
so the RNS benchmark is more likely to emerge than the mini-split benchmark
by a factor $\sim 3000$. 

\section{Conclusions}
\label{sec:conclude}

In this paper we examined the question: are finetuned or natural models
of particle physics more likely to emerge from
selection of string vacua within the multiverse?
We required as the anthropic condition that the atomic principle be
fulfilled in that the derived value of the weak scale lies within
the ABDS window.
We also assumed a scale invariant distribution for the superpotential
$\mu$ parameter in that for string theory with only a single mass scale,
the string scale, all other scales are equally likely.
We assumed the same distribution for $\mu_{SM}$.
For natural models with $\Delta_{EW}$ below say 30, then all
contributions to the weak scale lie within the ABDS window.
In this case, then a wide range of $\mu_{PU}$ values still lead to
$m_Z^{PU}$ within the ABDS window (since no finetuning of $\mu_{PU}$ is required).
For unnatural models with large contributions to the weak scale, then only
tiny ranges of $\mu_{PU}$ or $\mu_{SM}$ are allowed in order for $m_Z^{PU}$
to lie within the ABDS window.

Basically, particle physics models which require electroweak
finetuning may be possible on the landscape, but for a uniform distribution of
the tuning parameters, they are likely to be rare because the finetuned
solutions should be rare on the landscape relative to natural models.
This seems like common sense, but apparently contradicts the common
wisdom in the literature which asserts that the string landscape
provides motivation to take finetuned models as a plausible possibility
since the cosmological constant also appears to be finetuned.
The origin of the discrepancy may be traced to how the anthropic condition is
implemented. In many early works ({\it e.g.} \cite{Arkani-Hamed:2004ymt,Douglas:2004qg}), the anthropic penalty (finetuning factor)
is given as a factor $(m_{weak}/m_{soft})^2$ which favors $m_{soft}\sim m_{weak}$
but also allows for $m_{soft}\gg m_{weak}$.
This finetuning factor can be overwhelmed by a landscape draw to large soft
terms $m_{soft}^n$ with $n\ge 3$, thus favoring high scale SUSY breaking.
In contrast, we require as the anthropic condition that the derived value
of the weak scale lies within the ABDS window. In our method, if any contribution
to the weak scale in a given model is far beyond the measured value of
$m_{weak}$, then only a teensy range of $\mu_{PU}$ values are allowed to regain
the ABDS window. This is in contrast to natural models where a wide range
of $\mu_{PU}$ values lead to $m_{weak}$ within the ABDS window
(since no finetuning is needed).

\begin{figure}[!htbp]
\begin{center}
\includegraphics[height=0.3\textheight]{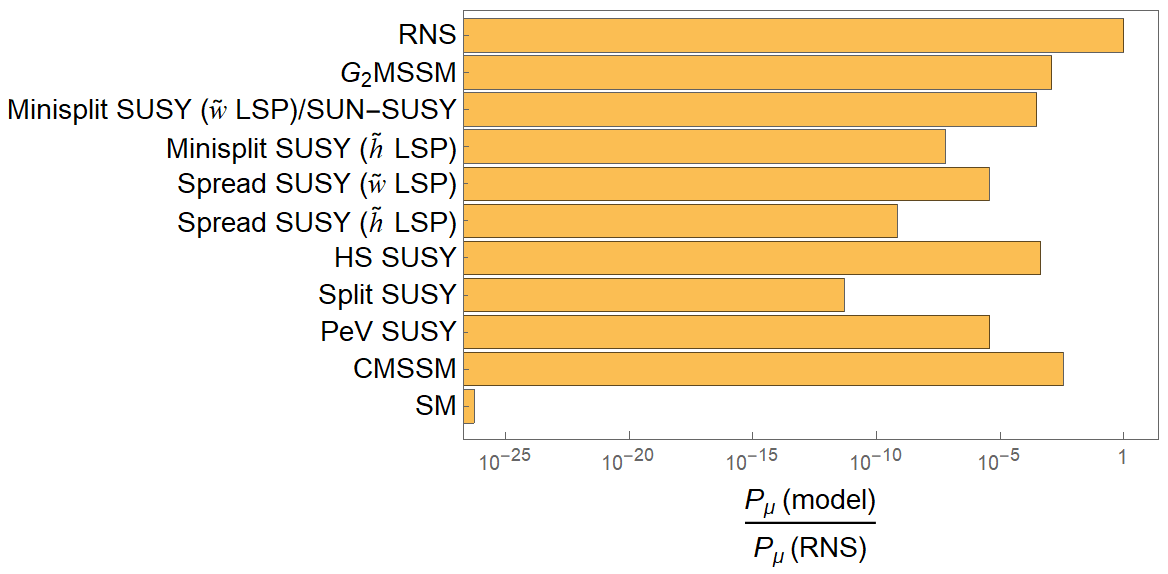}
\caption{Values of relative probabilities $P_\mu(model)/P_\mu(RNS)$
  for the various benchmark scenarios considered in the text.
\label{fig:Prob}}
\end{center}
\end{figure}
We examined the relative probabilities of various natural and finetuned SUSY
models, and the SM, to emerge from the landscape via a computable
measure of relative probabilities.
A summary of our results is shown in Fig. \ref{fig:Prob} where we
present a bar chart of the relative probabilities of the various
benchmark models considered here in relation to the radiatively-driven natural
SUSY benchmark where a wide range of $\mu_{PU}$ values lie within the
ABDS window.
Basically, the more finetuned a model is, the less likely it is to emerge
from the landscape in comparison to particle physics models with
electroweak naturalness.

\section*{Acknowledgements} 

This material is based upon work supported by the U.S. Department of Energy, 
Office of Science, Office of Basic Energy Sciences Energy Frontier Research 
Centers program under Award Number DE-SC-0009956 and U.S. Department of Energy 
Grant DE-SC-0017647. 




\bibliography{prob}

\begin{thebibliography}{10}
\expandafter\ifx\csname url\endcsname\relax
  \def\url#1{\texttt{#1}}\fi
\expandafter\ifx\csname urlprefix\endcsname\relax\def\urlprefix{URL }\fi
\expandafter\ifx\csname href\endcsname\relax
  \def\href#1#2{#2} \def\path#1{#1}\fi

\bibitem{Candelas:1985en}
P.~Candelas, G.~T. Horowitz, A.~Strominger, E.~Witten, {Vacuum Configurations
  for Superstrings}, Nucl. Phys. B 258 (1985) 46--74.
\newblock \href {https://doi.org/10.1016/0550-3213(85)90602-9}
  {\path{doi:10.1016/0550-3213(85)90602-9}}.

\bibitem{Acharya:2001gy}
B.~S. Acharya, E.~Witten, {Chiral fermions from manifolds of G(2) holonomy} (8
  2001).
\newblock \href {http://arxiv.org/abs/hep-th/0109152}
  {\path{arXiv:hep-th/0109152}}.

\bibitem{Acharya:2019mcu}
B.~S. Acharya, {Supersymmetry, Ricci Flat Manifolds and the String Landscape},
  JHEP 08 (2020) 128.
\newblock \href {http://arxiv.org/abs/1906.06886} {\path{arXiv:1906.06886}},
  \href {https://doi.org/10.1007/JHEP08(2020)128}
  {\path{doi:10.1007/JHEP08(2020)128}}.

\bibitem{GarciaEtxebarria:2020xsr}
I.~n. Garc\'\i{}a~Etxebarria, M.~Montero, K.~Sousa, I.~Valenzuela, {Nothing is
  certain in string compactifications}, JHEP 12 (2020) 032.
\newblock \href {http://arxiv.org/abs/2005.06494} {\path{arXiv:2005.06494}},
  \href {https://doi.org/10.1007/JHEP12(2020)032}
  {\path{doi:10.1007/JHEP12(2020)032}}.

\bibitem{Baer:2020kwz}
H.~Baer, V.~Barger, S.~Salam, D.~Sengupta, K.~Sinha, {Status of weak scale
  supersymmetry after LHC Run 2 and ton-scale noble liquid WIMP searches}, Eur.
  Phys. J. ST 229~(21) (2020) 3085--3141.
\newblock \href {http://arxiv.org/abs/2002.03013} {\path{arXiv:2002.03013}},
  \href {https://doi.org/10.1140/epjst/e2020-000020-x}
  {\path{doi:10.1140/epjst/e2020-000020-x}}.

\bibitem{Amaldi:1991cn}
U.~Amaldi, W.~de~Boer, H.~Furstenau, {Comparison of grand unified theories with
  electroweak and strong coupling constants measured at LEP}, Phys. Lett. B 260
  (1991) 447--455.
\newblock \href {https://doi.org/10.1016/0370-2693(91)91641-8}
  {\path{doi:10.1016/0370-2693(91)91641-8}}.

\bibitem{Dimopoulos:1981yj}
S.~Dimopoulos, S.~Raby, F.~Wilczek, {Supersymmetry and the Scale of
  Unification}, Phys. Rev. D 24 (1981) 1681--1683.
\newblock \href {https://doi.org/10.1103/PhysRevD.24.1681}
  {\path{doi:10.1103/PhysRevD.24.1681}}.

\bibitem{Dimopoulos:1981zb}
S.~Dimopoulos, H.~Georgi, {Softly Broken Supersymmetry and SU(5)}, Nucl. Phys.
  B 193 (1981) 150--162.
\newblock \href {https://doi.org/10.1016/0550-3213(81)90522-8}
  {\path{doi:10.1016/0550-3213(81)90522-8}}.

\bibitem{Slavich:2020zjv}
P.~Slavich, et~al., {Higgs-mass predictions in the MSSM and beyond}, Eur. Phys.
  J. C 81~(5) (2021) 450.
\newblock \href {http://arxiv.org/abs/2012.15629} {\path{arXiv:2012.15629}},
  \href {https://doi.org/10.1140/epjc/s10052-021-09198-2}
  {\path{doi:10.1140/epjc/s10052-021-09198-2}}.

\bibitem{Dine:2015xga}
M.~Dine, {Naturalness Under Stress}, Ann. Rev. Nucl. Part. Sci. 65 (2015)
  43--62.
\newblock \href {http://arxiv.org/abs/1501.01035} {\path{arXiv:1501.01035}},
  \href {https://doi.org/10.1146/annurev-nucl-102014-022053}
  {\path{doi:10.1146/annurev-nucl-102014-022053}}.

\bibitem{Susskind:2003kw}
L.~Susskind, {The Anthropic landscape of string theory} (2003) 247--266\href
  {http://arxiv.org/abs/hep-th/0302219} {\path{arXiv:hep-th/0302219}}.

\bibitem{Douglas:2006es}
M.~R. Douglas, S.~Kachru, {Flux compactification}, Rev. Mod. Phys. 79 (2007)
  733--796.
\newblock \href {http://arxiv.org/abs/hep-th/0610102}
  {\path{arXiv:hep-th/0610102}}, \href
  {https://doi.org/10.1103/RevModPhys.79.733}
  {\path{doi:10.1103/RevModPhys.79.733}}.

\bibitem{Bousso:2000xa}
R.~Bousso, J.~Polchinski, {Quantization of four form fluxes and dynamical
  neutralization of the cosmological constant}, JHEP 06 (2000) 006.
\newblock \href {http://arxiv.org/abs/hep-th/0004134}
  {\path{arXiv:hep-th/0004134}}, \href
  {https://doi.org/10.1088/1126-6708/2000/06/006}
  {\path{doi:10.1088/1126-6708/2000/06/006}}.

\bibitem{Ashok:2003gk}
S.~Ashok, M.~R. Douglas, {Counting flux vacua}, JHEP 01 (2004) 060.
\newblock \href {http://arxiv.org/abs/hep-th/0307049}
  {\path{arXiv:hep-th/0307049}}, \href
  {https://doi.org/10.1088/1126-6708/2004/01/060}
  {\path{doi:10.1088/1126-6708/2004/01/060}}.

\bibitem{Linde:2015edk}
A.~Linde, {A brief history of the multiverse}, Rept. Prog. Phys. 80~(2) (2017)
  022001.
\newblock \href {http://arxiv.org/abs/1512.01203} {\path{arXiv:1512.01203}},
  \href {https://doi.org/10.1088/1361-6633/aa50e4}
  {\path{doi:10.1088/1361-6633/aa50e4}}.

\bibitem{Weinberg:1987dv}
S.~Weinberg, {Anthropic Bound on the Cosmological Constant}, Phys. Rev. Lett.
  59 (1987) 2607.
\newblock \href {https://doi.org/10.1103/PhysRevLett.59.2607}
  {\path{doi:10.1103/PhysRevLett.59.2607}}.

\bibitem{Agrawal:1997gf}
V.~Agrawal, S.~M. Barr, J.~F. Donoghue, D.~Seckel, {Viable range of the mass
  scale of the standard model}, Phys. Rev. D 57 (1998) 5480--5492.
\newblock \href {http://arxiv.org/abs/hep-ph/9707380}
  {\path{arXiv:hep-ph/9707380}}, \href
  {https://doi.org/10.1103/PhysRevD.57.5480}
  {\path{doi:10.1103/PhysRevD.57.5480}}.

\bibitem{Agrawal:1998xa}
V.~Agrawal, S.~M. Barr, J.~F. Donoghue, D.~Seckel, {Anthropic considerations in
  multiple domain theories and the scale of electroweak symmetry breaking},
  Phys. Rev. Lett. 80 (1998) 1822--1825.
\newblock \href {http://arxiv.org/abs/hep-ph/9801253}
  {\path{arXiv:hep-ph/9801253}}, \href
  {https://doi.org/10.1103/PhysRevLett.80.1822}
  {\path{doi:10.1103/PhysRevLett.80.1822}}.

\bibitem{Arkani-Hamed:2004ymt}
N.~Arkani-Hamed, S.~Dimopoulos, {Supersymmetric unification without low energy
  supersymmetry and signatures for fine-tuning at the LHC}, JHEP 06 (2005) 073.
\newblock \href {http://arxiv.org/abs/hep-th/0405159}
  {\path{arXiv:hep-th/0405159}}, \href
  {https://doi.org/10.1088/1126-6708/2005/06/073}
  {\path{doi:10.1088/1126-6708/2005/06/073}}.

\bibitem{Giudice:2004tc}
G.~F. Giudice, A.~Romanino, {Split supersymmetry}, Nucl. Phys. B 699 (2004)
  65--89, [Erratum: Nucl.Phys.B 706, 487--487 (2005)].
\newblock \href {http://arxiv.org/abs/hep-ph/0406088}
  {\path{arXiv:hep-ph/0406088}}, \href
  {https://doi.org/10.1016/j.nuclphysb.2004.08.001}
  {\path{doi:10.1016/j.nuclphysb.2004.08.001}}.

\bibitem{Arkani-Hamed:2004zhs}
N.~Arkani-Hamed, S.~Dimopoulos, G.~F. Giudice, A.~Romanino, {Aspects of split
  supersymmetry}, Nucl. Phys. B 709 (2005) 3--46.
\newblock \href {http://arxiv.org/abs/hep-ph/0409232}
  {\path{arXiv:hep-ph/0409232}}, \href
  {https://doi.org/10.1016/j.nuclphysb.2004.12.026}
  {\path{doi:10.1016/j.nuclphysb.2004.12.026}}.

\bibitem{Dine:1990jd}
M.~Dine, A.~Kagan, S.~Samuel, {Naturalness in Supersymmetry, or Raising the
  Supersymmetry Breaking Scale}, Phys. Lett. B 243 (1990) 250--256.
\newblock \href {https://doi.org/10.1016/0370-2693(90)90847-Y}
  {\path{doi:10.1016/0370-2693(90)90847-Y}}.

\bibitem{Cohen:1996vb}
A.~G. Cohen, D.~B. Kaplan, A.~E. Nelson, {The More minimal supersymmetric
  standard model}, Phys. Lett. B 388 (1996) 588--598.
\newblock \href {http://arxiv.org/abs/hep-ph/9607394}
  {\path{arXiv:hep-ph/9607394}}, \href
  {https://doi.org/10.1016/S0370-2693(96)01183-5}
  {\path{doi:10.1016/S0370-2693(96)01183-5}}.

\bibitem{Bagger:1999sy}
J.~A. Bagger, J.~L. Feng, N.~Polonsky, R.-J. Zhang, {Superheavy supersymmetry
  from scalar mass: A parameter fixed points}, Phys. Lett. B 473 (2000)
  264--271.
\newblock \href {http://arxiv.org/abs/hep-ph/9911255}
  {\path{arXiv:hep-ph/9911255}}, \href
  {https://doi.org/10.1016/S0370-2693(99)01501-4}
  {\path{doi:10.1016/S0370-2693(99)01501-4}}.

\bibitem{Bae:2022okh}
K.~J. Bae, H.~Baer, V.~Barger, R.~W. Deal, {The cosmological moduli problem and
  naturalness}, JHEP 02 (2022) 138.
\newblock \href {http://arxiv.org/abs/2201.06633} {\path{arXiv:2201.06633}},
  \href {https://doi.org/10.1007/JHEP02(2022)138}
  {\path{doi:10.1007/JHEP02(2022)138}}.

\bibitem{Hall:2011jd}
L.~J. Hall, Y.~Nomura, {Spread Supersymmetry}, JHEP 01 (2012) 082.
\newblock \href {http://arxiv.org/abs/1111.4519} {\path{arXiv:1111.4519}},
  \href {https://doi.org/10.1007/JHEP01(2012)082}
  {\path{doi:10.1007/JHEP01(2012)082}}.

\bibitem{Arvanitaki:2012ps}
A.~Arvanitaki, N.~Craig, S.~Dimopoulos, G.~Villadoro, {Mini-Split}, JHEP 02
  (2013) 126.
\newblock \href {http://arxiv.org/abs/1210.0555} {\path{arXiv:1210.0555}},
  \href {https://doi.org/10.1007/JHEP02(2013)126}
  {\path{doi:10.1007/JHEP02(2013)126}}.

\bibitem{Douglas:2004qg}
M.~R. Douglas, {Statistical analysis of the supersymmetry breaking scale} (5
  2004).
\newblock \href {http://arxiv.org/abs/hep-th/0405279}
  {\path{arXiv:hep-th/0405279}}.

\bibitem{Susskind:2004uv}
L.~Susskind, {Supersymmetry breaking in the anthropic landscape}, in: {From
  Fields to Strings: Circumnavigating Theoretical Physics: A Conference in
  Tribute to Ian Kogan}, 2004, pp. 1745--1749.
\newblock \href {http://arxiv.org/abs/hep-th/0405189}
  {\path{arXiv:hep-th/0405189}}, \href
  {https://doi.org/10.1142/9789812775344_0040}
  {\path{doi:10.1142/9789812775344_0040}}.

\bibitem{Arkani-Hamed:2005zuc}
N.~Arkani-Hamed, S.~Dimopoulos, S.~Kachru, {Predictive landscapes and new
  physics at a TeV} (1 2005).
\newblock \href {http://arxiv.org/abs/hep-th/0501082}
  {\path{arXiv:hep-th/0501082}}.

\bibitem{Baer:2016lpj}
H.~Baer, V.~Barger, M.~Savoy, H.~Serce, {The Higgs mass and natural
  supersymmetric spectrum from the landscape}, Phys. Lett. B 758 (2016)
  113--117.
\newblock \href {http://arxiv.org/abs/1602.07697} {\path{arXiv:1602.07697}},
  \href {https://doi.org/10.1016/j.physletb.2016.05.010}
  {\path{doi:10.1016/j.physletb.2016.05.010}}.

\bibitem{Baer:2017uvn}
H.~Baer, V.~Barger, H.~Serce, K.~Sinha, {Higgs and superparticle mass
  predictions from the landscape}, JHEP 03 (2018) 002.
\newblock \href {http://arxiv.org/abs/1712.01399} {\path{arXiv:1712.01399}},
  \href {https://doi.org/10.1007/JHEP03(2018)002}
  {\path{doi:10.1007/JHEP03(2018)002}}.

\bibitem{Baer:2015rja}
H.~Baer, V.~Barger, M.~Savoy, {Upper bounds on sparticle masses from
  naturalness or how to disprove weak scale supersymmetry}, Phys. Rev. D 93~(3)
  (2016) 035016.
\newblock \href {http://arxiv.org/abs/1509.02929} {\path{arXiv:1509.02929}},
  \href {https://doi.org/10.1103/PhysRevD.93.035016}
  {\path{doi:10.1103/PhysRevD.93.035016}}.

\bibitem{Giudice:2008bi}
G.~F. Giudice, {Naturally Speaking: The Naturalness Criterion and Physics at
  the LHC} (2008) 155--178\href {http://arxiv.org/abs/0801.2562}
  {\path{arXiv:0801.2562}}, \href {https://doi.org/10.1142/9789812779762_0010}
  {\path{doi:10.1142/9789812779762_0010}}.

\bibitem{Baer:2012cf}
H.~Baer, V.~Barger, P.~Huang, D.~Mickelson, A.~Mustafayev, X.~Tata, {Radiative
  natural supersymmetry: Reconciling electroweak fine-tuning and the Higgs
  boson mass}, Phys. Rev. D 87~(11) (2013) 115028.
\newblock \href {http://arxiv.org/abs/1212.2655} {\path{arXiv:1212.2655}},
  \href {https://doi.org/10.1103/PhysRevD.87.115028}
  {\path{doi:10.1103/PhysRevD.87.115028}}.

\bibitem{Baer:2021tta}
H.~Baer, V.~Barger, D.~Martinez, {Comparison of SUSY spectra generators for
  natural SUSY and string landscape predictions}, Eur. Phys. J. C 82~(2) (2022)
  172.
\newblock \href {http://arxiv.org/abs/2111.03096} {\path{arXiv:2111.03096}},
  \href {https://doi.org/10.1140/epjc/s10052-022-10141-2}
  {\path{doi:10.1140/epjc/s10052-022-10141-2}}.

\bibitem{Dedes:2002dy}
A.~Dedes, P.~Slavich, {Two loop corrections to radiative electroweak symmetry
  breaking in the MSSM}, Nucl. Phys. B 657 (2003) 333--354.
\newblock \href {http://arxiv.org/abs/hep-ph/0212132}
  {\path{arXiv:hep-ph/0212132}}, \href
  {https://doi.org/10.1016/S0550-3213(03)00173-1}
  {\path{doi:10.1016/S0550-3213(03)00173-1}}.

\bibitem{Baer:2012up}
H.~Baer, V.~Barger, P.~Huang, A.~Mustafayev, X.~Tata, {Radiative natural SUSY
  with a 125 GeV Higgs boson}, Phys. Rev. Lett. 109 (2012) 161802.
\newblock \href {http://arxiv.org/abs/1207.3343} {\path{arXiv:1207.3343}},
  \href {https://doi.org/10.1103/PhysRevLett.109.161802}
  {\path{doi:10.1103/PhysRevLett.109.161802}}.

\bibitem{Chamseddine:2000nk}
A.~H. Chamseddine, R.~L. Arnowitt, P.~Nath, {Supergravity unification}, Nucl.
  Phys. B Proc. Suppl. 101 (2001) 145--153.
\newblock \href {http://arxiv.org/abs/hep-ph/0102286}
  {\path{arXiv:hep-ph/0102286}}, \href
  {https://doi.org/10.1016/S0920-5632(01)01501-8}
  {\path{doi:10.1016/S0920-5632(01)01501-8}}.

\bibitem{Kane:1993td}
G.~L. Kane, C.~F. Kolda, L.~Roszkowski, J.~D. Wells, {Study of constrained
  minimal supersymmetry}, Phys. Rev. D 49 (1994) 6173--6210.
\newblock \href {http://arxiv.org/abs/hep-ph/9312272}
  {\path{arXiv:hep-ph/9312272}}, \href
  {https://doi.org/10.1103/PhysRevD.49.6173}
  {\path{doi:10.1103/PhysRevD.49.6173}}.

\bibitem{Baer:2012mv}
H.~Baer, V.~Barger, P.~Huang, D.~Mickelson, A.~Mustafayev, X.~Tata, {Post-LHC7
  fine-tuning in the minimal supergravity/CMSSM model with a 125 GeV Higgs
  boson}, Phys. Rev. D 87~(3) (2013) 035017.
\newblock \href {http://arxiv.org/abs/1210.3019} {\path{arXiv:1210.3019}},
  \href {https://doi.org/10.1103/PhysRevD.87.035017}
  {\path{doi:10.1103/PhysRevD.87.035017}}.

\bibitem{Baer:2014ica}
H.~Baer, V.~Barger, D.~Mickelson, M.~Padeffke-Kirkland, {SUSY models under
  siege: LHC constraints and electroweak fine-tuning}, Phys. Rev. D 89~(11)
  (2014) 115019.
\newblock \href {http://arxiv.org/abs/1404.2277} {\path{arXiv:1404.2277}},
  \href {https://doi.org/10.1103/PhysRevD.89.115019}
  {\path{doi:10.1103/PhysRevD.89.115019}}.

\bibitem{Paige:2003mg}
F.~E. Paige, S.~D. Protopopescu, H.~Baer, X.~Tata, {ISAJET 7.69: A Monte Carlo
  event generator for pp, anti-p p, and e+e- reactions} (12 2003).
\newblock \href {http://arxiv.org/abs/hep-ph/0312045}
  {\path{arXiv:hep-ph/0312045}}.

\bibitem{Baer:1994nc}
H.~Baer, C.-H. Chen, R.~B. Munroe, F.~E. Paige, X.~Tata, {Multichannel search
  for minimal supergravity at $p \bar{p}$ and $e^{+} e^{-}$ colliders}, Phys.
  Rev. D 51 (1995) 1046--1050.
\newblock \href {http://arxiv.org/abs/hep-ph/9408265}
  {\path{arXiv:hep-ph/9408265}}, \href
  {https://doi.org/10.1103/PhysRevD.51.1046}
  {\path{doi:10.1103/PhysRevD.51.1046}}.

\bibitem{Baer:2014eja}
H.~Baer, K.-Y. Choi, J.~E. Kim, L.~Roszkowski, {Dark matter production in the
  early Universe: beyond the thermal WIMP paradigm}, Phys. Rept. 555 (2015)
  1--60.
\newblock \href {http://arxiv.org/abs/1407.0017} {\path{arXiv:1407.0017}},
  \href {https://doi.org/10.1016/j.physrep.2014.10.002}
  {\path{doi:10.1016/j.physrep.2014.10.002}}.

\bibitem{Wells:2003tf}
J.~D. Wells, {Implications of supersymmetry breaking with a little hierarchy
  between gauginos and scalars}, in: {11th International Conference on
  Supersymmetry and the Unification of Fundamental Interactions}, 2003.
\newblock \href {http://arxiv.org/abs/hep-ph/0306127}
  {\path{arXiv:hep-ph/0306127}}.

\bibitem{Wells:2004di}
J.~D. Wells, {PeV-scale supersymmetry}, Phys. Rev. D 71 (2005) 015013.
\newblock \href {http://arxiv.org/abs/hep-ph/0411041}
  {\path{arXiv:hep-ph/0411041}}, \href
  {https://doi.org/10.1103/PhysRevD.71.015013}
  {\path{doi:10.1103/PhysRevD.71.015013}}.

\bibitem{Giudice:2011cg}
G.~F. Giudice, A.~Strumia, {Probing High-Scale and Split Supersymmetry with
  Higgs Mass Measurements}, Nucl. Phys. B 858 (2012) 63--83.
\newblock \href {http://arxiv.org/abs/1108.6077} {\path{arXiv:1108.6077}},
  \href {https://doi.org/10.1016/j.nuclphysb.2012.01.001}
  {\path{doi:10.1016/j.nuclphysb.2012.01.001}}.

\bibitem{Barger:2005qy}
V.~Barger, J.~Jiang, P.~Langacker, T.~Li, {Non-canonical gauge coupling
  unification in high-scale supersymmetry breaking}, Nucl. Phys. B 726 (2005)
  149--170.
\newblock \href {http://arxiv.org/abs/hep-ph/0504093}
  {\path{arXiv:hep-ph/0504093}}, \href
  {https://doi.org/10.1016/j.nuclphysb.2005.08.007}
  {\path{doi:10.1016/j.nuclphysb.2005.08.007}}.

\bibitem{Barger:2007qb}
V.~Barger, N.~G. Deshpande, J.~Jiang, P.~Langacker, T.~Li, {Implications of
  Canonical Gauge Coupling Unification in High-Scale Supersymmetry Breaking},
  Nucl. Phys. B 793 (2008) 307--325.
\newblock \href {http://arxiv.org/abs/hep-ph/0701136}
  {\path{arXiv:hep-ph/0701136}}, \href
  {https://doi.org/10.1016/j.nuclphysb.2007.10.013}
  {\path{doi:10.1016/j.nuclphysb.2007.10.013}}.

\bibitem{Ellis:2017erg}
S.~A.~R. Ellis, J.~D. Wells, {High-scale supersymmetry, the Higgs boson mass,
  and gauge unification}, Phys. Rev. D 96~(5) (2017) 055024.
\newblock \href {http://arxiv.org/abs/1706.00013} {\path{arXiv:1706.00013}},
  \href {https://doi.org/10.1103/PhysRevD.96.055024}
  {\path{doi:10.1103/PhysRevD.96.055024}}.

\bibitem{Bagnaschi:2014rsa}
E.~Bagnaschi, G.~F. Giudice, P.~Slavich, A.~Strumia, {Higgs Mass and Unnatural
  Supersymmetry}, JHEP 09 (2014) 092.
\newblock \href {http://arxiv.org/abs/1407.4081} {\path{arXiv:1407.4081}},
  \href {https://doi.org/10.1007/JHEP09(2014)092}
  {\path{doi:10.1007/JHEP09(2014)092}}.

\bibitem{Arkani-Hamed:2012fhg}
N.~Arkani-Hamed, A.~Gupta, D.~E. Kaplan, N.~Weiner, T.~Zorawski, {Simply
  Unnatural Supersymmetry} (12 2012).
\newblock \href {http://arxiv.org/abs/1212.6971} {\path{arXiv:1212.6971}}.

\bibitem{Tegmark:2005dy}
M.~Tegmark, A.~Aguirre, M.~Rees, F.~Wilczek, {Dimensionless constants,
  cosmology and other dark matters}, Phys. Rev. D 73 (2006) 023505.
\newblock \href {http://arxiv.org/abs/astro-ph/0511774}
  {\path{arXiv:astro-ph/0511774}}, \href
  {https://doi.org/10.1103/PhysRevD.73.023505}
  {\path{doi:10.1103/PhysRevD.73.023505}}.

\bibitem{Acharya:2006ia}
B.~S. Acharya, K.~Bobkov, G.~Kane, P.~Kumar, D.~Vaman, {An M theory Solution to
  the Hierarchy Problem}, Phys. Rev. Lett. 97 (2006) 191601.
\newblock \href {http://arxiv.org/abs/hep-th/0606262}
  {\path{arXiv:hep-th/0606262}}, \href
  {https://doi.org/10.1103/PhysRevLett.97.191601}
  {\path{doi:10.1103/PhysRevLett.97.191601}}.

\bibitem{Acharya:2008zi}
B.~S. Acharya, K.~Bobkov, G.~L. Kane, J.~Shao, P.~Kumar, {The G(2)-MSSM: An M
  Theory motivated model of Particle Physics}, Phys. Rev. D 78 (2008) 065038.
\newblock \href {http://arxiv.org/abs/0801.0478} {\path{arXiv:0801.0478}},
  \href {https://doi.org/10.1103/PhysRevD.78.065038}
  {\path{doi:10.1103/PhysRevD.78.065038}}.

\bibitem{Acharya:2008bk}
B.~S. Acharya, P.~Kumar, K.~Bobkov, G.~Kane, J.~Shao, S.~Watson, {Non-thermal
  Dark Matter and the Moduli Problem in String Frameworks}, JHEP 06 (2008) 064.
\newblock \href {http://arxiv.org/abs/0804.0863} {\path{arXiv:0804.0863}},
  \href {https://doi.org/10.1088/1126-6708/2008/06/064}
  {\path{doi:10.1088/1126-6708/2008/06/064}}.

\bibitem{Acharya:2016fge}
B.~S. Acharya, S.~A.~R. Ellis, G.~L. Kane, B.~D. Nelson, M.~J. Perry, {The
  lightest visible-sector supersymmetric particle is likely to be unstable},
  Phys. Rev. Lett. 117 (2016) 181802.
\newblock \href {http://arxiv.org/abs/1604.05320} {\path{arXiv:1604.05320}},
  \href {https://doi.org/10.1103/PhysRevLett.117.181802}
  {\path{doi:10.1103/PhysRevLett.117.181802}}.

\bibitem{Baer:2016hfa}
H.~Baer, V.~Barger, H.~Serce, X.~Tata, {Natural generalized mirage mediation},
  Phys. Rev. D 94~(11) (2016) 115017.
\newblock \href {http://arxiv.org/abs/1610.06205} {\path{arXiv:1610.06205}},
  \href {https://doi.org/10.1103/PhysRevD.94.115017}
  {\path{doi:10.1103/PhysRevD.94.115017}}.

\bibitem{Baer:2018hwa}
H.~Baer, V.~Barger, D.~Sengupta, {Anomaly mediated SUSY breaking model
  retrofitted for naturalness}, Phys. Rev. D 98~(1) (2018) 015039.
\newblock \href {http://arxiv.org/abs/1801.09730} {\path{arXiv:1801.09730}},
  \href {https://doi.org/10.1103/PhysRevD.98.015039}
  {\path{doi:10.1103/PhysRevD.98.015039}}.

\bibitem{Broeckel:2020fdz}
I.~Broeckel, M.~Cicoli, A.~Maharana, K.~Singh, K.~Sinha, {Moduli Stabilisation
  and the Statistics of SUSY Breaking in the Landscape}, JHEP 10 (2020) 015.
\newblock \href {http://arxiv.org/abs/2007.04327} {\path{arXiv:2007.04327}},
  \href {https://doi.org/10.1007/JHEP09(2021)019}
  {\path{doi:10.1007/JHEP09(2021)019}}.

\bibitem{Kachru:2003aw}
S.~Kachru, R.~Kallosh, A.~D. Linde, S.~P. Trivedi, {De Sitter vacua in string
  theory}, Phys. Rev. D 68 (2003) 046005.
\newblock \href {http://arxiv.org/abs/hep-th/0301240}
  {\path{arXiv:hep-th/0301240}}, \href
  {https://doi.org/10.1103/PhysRevD.68.046005}
  {\path{doi:10.1103/PhysRevD.68.046005}}.

\bibitem{Baer:2011ab}
H.~Baer, V.~Barger, A.~Mustafayev, {Implications of a 125 GeV Higgs scalar for
  LHC SUSY and neutralino dark matter searches}, Phys. Rev. D 85 (2012) 075010.
\newblock \href {http://arxiv.org/abs/1112.3017} {\path{arXiv:1112.3017}},
  \href {https://doi.org/10.1103/PhysRevD.85.075010}
  {\path{doi:10.1103/PhysRevD.85.075010}}.

\bibitem{Douglas:2004zg}
M.~R. Douglas, {Basic results in vacuum statistics}, Comptes Rendus Physique 5
  (2004) 965--977.
\newblock \href {http://arxiv.org/abs/hep-th/0409207}
  {\path{arXiv:hep-th/0409207}}, \href
  {https://doi.org/10.1016/j.crhy.2004.09.008}
  {\path{doi:10.1016/j.crhy.2004.09.008}}.

\bibitem{Baer:2019cae}
H.~Baer, V.~Barger, S.~Salam, {Naturalness versus stringy naturalness (with
  implications for collider and dark matter searches}, Phys. Rev. Research. 1
  (2019) 023001.
\newblock \href {http://arxiv.org/abs/1906.07741} {\path{arXiv:1906.07741}},
  \href {https://doi.org/10.1103/PhysRevResearch.1.023001}
  {\path{doi:10.1103/PhysRevResearch.1.023001}}.

\bibitem{Baer:2022wxe}
H.~Baer, V.~Barger, D.~Martinez, S.~Salam, {Radiative natural supersymmetry
  emergent from the string landscape}, JHEP 03 (2022) 186.
\newblock \href {http://arxiv.org/abs/2202.07046} {\path{arXiv:2202.07046}},
  \href {https://doi.org/10.1007/JHEP03(2022)186}
  {\path{doi:10.1007/JHEP03(2022)186}}.

\bibitem{Cremmer:1982en}
E.~Cremmer, S.~Ferrara, L.~Girardello, A.~Van~Proeyen, {Yang-Mills Theories
  with Local Supersymmetry: Lagrangian, Transformation Laws and SuperHiggs
  Effect}, Nucl. Phys. B 212 (1983) 413.
\newblock \href {https://doi.org/10.1016/0550-3213(83)90679-X}
  {\path{doi:10.1016/0550-3213(83)90679-X}}.

\bibitem{Soni:1983rm}
S.~K. Soni, H.~A. Weldon, {Analysis of the Supersymmetry Breaking Induced by
  N=1 Supergravity Theories}, Phys. Lett. B 126 (1983) 215--219.
\newblock \href {https://doi.org/10.1016/0370-2693(83)90593-2}
  {\path{doi:10.1016/0370-2693(83)90593-2}}.

\bibitem{Kaplunovsky:1993rd}
V.~S. Kaplunovsky, J.~Louis, {Model independent analysis of soft terms in
  effective supergravity and in string theory}, Phys. Lett. B 306 (1993)
  269--275.
\newblock \href {http://arxiv.org/abs/hep-th/9303040}
  {\path{arXiv:hep-th/9303040}}, \href
  {https://doi.org/10.1016/0370-2693(93)90078-V}
  {\path{doi:10.1016/0370-2693(93)90078-V}}.

\bibitem{Brignole:1993dj}
A.~Brignole, L.~E. Ibanez, C.~Munoz, {Towards a theory of soft terms for the
  supersymmetric Standard Model}, Nucl. Phys. B 422 (1994) 125--171, [Erratum:
  Nucl.Phys.B 436, 747--748 (1995)].
\newblock \href {http://arxiv.org/abs/hep-ph/9308271}
  {\path{arXiv:hep-ph/9308271}}, \href
  {https://doi.org/10.1016/0550-3213(94)00068-9}
  {\path{doi:10.1016/0550-3213(94)00068-9}}.

\bibitem{Baer:2020vad}
H.~Baer, V.~Barger, S.~Salam, D.~Sengupta, {String landscape guide to soft SUSY
  breaking terms}, Phys. Rev. D 102~(7) (2020) 075012.
\newblock \href {http://arxiv.org/abs/2005.13577} {\path{arXiv:2005.13577}},
  \href {https://doi.org/10.1103/PhysRevD.102.075012}
  {\path{doi:10.1103/PhysRevD.102.075012}}.

\bibitem{Kim:1983dt}
J.~E. Kim, H.~P. Nilles, {The mu Problem and the Strong CP Problem}, Phys.
  Lett. B 138 (1984) 150--154.
\newblock \href {https://doi.org/10.1016/0370-2693(84)91890-2}
  {\path{doi:10.1016/0370-2693(84)91890-2}}.

\bibitem{Bae:2019dgg}
K.~J. Bae, H.~Baer, V.~Barger, D.~Sengupta, {Revisiting the SUSY $\mu$ problem
  and its solutions in the LHC era}, Phys. Rev. D 99~(11) (2019) 115027.
\newblock \href {http://arxiv.org/abs/1902.10748} {\path{arXiv:1902.10748}},
  \href {https://doi.org/10.1103/PhysRevD.99.115027}
  {\path{doi:10.1103/PhysRevD.99.115027}}.

\bibitem{Baer:2021vrk}
H.~Baer, V.~Barger, D.~Sengupta, R.~W. Deal, {Distribution of supersymmetry
  \ensuremath{\mu} parameter and Peccei-Quinn scale fa from the landscape},
  Phys. Rev. D 104~(1) (2021) 015037.
\newblock \href {http://arxiv.org/abs/2104.03803} {\path{arXiv:2104.03803}},
  \href {https://doi.org/10.1103/PhysRevD.104.015037}
  {\path{doi:10.1103/PhysRevD.104.015037}}.

\bibitem{Donoghue:2005cf}
J.~F. Donoghue, K.~Dutta, A.~Ross, {Quark and lepton masses and mixing in the
  landscape}, Phys. Rev. D 73 (2006) 113002.
\newblock \href {http://arxiv.org/abs/hep-ph/0511219}
  {\path{arXiv:hep-ph/0511219}}, \href
  {https://doi.org/10.1103/PhysRevD.73.113002}
  {\path{doi:10.1103/PhysRevD.73.113002}}.

\bibitem{Degrassi:2012ry}
G.~Degrassi, S.~Di~Vita, J.~Elias-Miro, J.~R. Espinosa, G.~F. Giudice,
  G.~Isidori, A.~Strumia, {Higgs mass and vacuum stability in the Standard
  Model at NNLO}, JHEP 08 (2012) 098.
\newblock \href {http://arxiv.org/abs/1205.6497} {\path{arXiv:1205.6497}},
  \href {https://doi.org/10.1007/JHEP08(2012)098}
  {\path{doi:10.1007/JHEP08(2012)098}}.

\bibitem{Athron:2014yba}
P.~Athron, J.-h. Park, D.~St\"ockinger, A.~Voigt, {FlexibleSUSY\textemdash{}A
  spectrum generator generator for supersymmetric models}, Comput. Phys.
  Commun. 190 (2015) 139--172.
\newblock \href {http://arxiv.org/abs/1406.2319} {\path{arXiv:1406.2319}},
  \href {https://doi.org/10.1016/j.cpc.2014.12.020}
  {\path{doi:10.1016/j.cpc.2014.12.020}}.

\bibitem{Athron:2016fuq}
P.~Athron, J.-h. Park, T.~Steudtner, D.~St\"ockinger, A.~Voigt, {Precise Higgs
  mass calculations in (non-)minimal supersymmetry at both high and low
  scales}, JHEP 01 (2017) 079.
\newblock \href {http://arxiv.org/abs/1609.00371} {\path{arXiv:1609.00371}},
  \href {https://doi.org/10.1007/JHEP01(2017)079}
  {\path{doi:10.1007/JHEP01(2017)079}}.

\bibitem{Athron:2017fvs}
P.~Athron, M.~Bach, D.~Harries, T.~Kwasnitza, J.-h. Park, D.~St\"ockinger,
  A.~Voigt, J.~Ziebell, {FlexibleSUSY 2.0: Extensions to investigate the
  phenomenology of SUSY and non-SUSY models}, Comput. Phys. Commun. 230 (2018)
  145--217.
\newblock \href {http://arxiv.org/abs/1710.03760} {\path{arXiv:1710.03760}},
  \href {https://doi.org/10.1016/j.cpc.2018.04.016}
  {\path{doi:10.1016/j.cpc.2018.04.016}}.

\end{thebibliography}
\bibliographystyle{elsarticle-num}

\end{document}